\documentclass[journal ]{new-aiaa}

\usepackage[utf8]{inputenc}
\usepackage{textcomp}
\usepackage{graphicx}
\usepackage{amsmath}
\usepackage[version=4]{mhchem}
\usepackage{siunitx}
\usepackage{mathtools}
\usepackage{longtable,tabularx}
\usepackage{nomencl}
\usepackage{mathrsfs} 
\usepackage{mathtools} 
\usepackage{optidef}
\usepackage{multicol}
\usepackage{svg}
\usepackage{lipsum}
\usepackage{calrsfs}
\usepackage{scalerel}
\usepackage{stackengine}
\usepackage{booktabs}
\usepackage{caption}
\usepackage{enumerate}
\usepackage{comment}
\usepackage{rotating}
\usepackage[english]{babel}

\newtheorem{definition}{Definition}
\newtheorem{theorem}{Theorem}
\newtheorem{remark}{Remark}
\usepackage{comment}

\newcommand{\tends}{\rightarrow}

\title{Quadratic Programming Approach to Flight Envelope Protection Using Control Barrier Functions}

\author{Johannes Autenrieb\footnote{Research Scientist, Department of Flight Dynamics and Simulation, email: johannes.autenrieb@dlr.de}}
\affil{German Aerospace Center (DLR), Institute of Flight Systems, 38108, Braunschweig, Germany}

\begin{document}

\maketitle

\begin{abstract}
Ensuring the safe operation of aerospace systems within their prescribed flight envelope is a fundamental requirement for modern flight control systems. Flight envelope protection (FEP) prevents violations of aerodynamic, structural, and performance constraints, mitigating risks such as stall, excessive loads, and loss of control. Conventional FEP approaches, such as reference clipping via saturation functions and model-based command filtering, impose constraints at the reference input level but often fail to account for closed-loop system dynamics, potentially leading to constraint violations during transients. This paper introduces a new approach to flight envelope protection by employing a quadratic-programming-based safety filter using control barrier functions to dynamically enforce flight envelope constraints while preserving control performance. Unlike traditional reference filtering methods, the proposed control barrier function-based safety filter actively ensures forward invariance of the safe flight envelope set while seamlessly integrating with existing control architectures. The framework is implemented in a nonlinear missile flight control system and evaluated in a simulated environment. The results demonstrate its ability to prevent constraint violations while minimizing conservatism, offering a robust alternative to existing flight envelope protection methodologies.
\end{abstract}

\section{Introduction}
\label{sec:Introduction}
In recent years, the aerospace field has undergone significant development, with new configurations and expanding fields of application emerging across both the civil and defense sectors. Modern aerospace vehicles—such as high-speed missiles, hypersonic glide vehicles, and next-generation autonomous aircraft—push the limits of aerodynamics, propulsion, and structural integrity while operating at extreme speeds, dynamic pressures, and control authority limits. Modern flight control systems must incorporate verifiable control strategies that ensure high performance while providing theoretical guarantees of safe operation within the available flight envelope.\\

Flight envelope protection (FEP) plays a crucial role in this context, ensuring that aerospace vehicles remain within safe operational limits defined by aerodynamic, structural, and performance constraints. The flight envelope encompasses critical parameters, including speed, load factor, altitude, and angle of attack, all of which must be respected to prevent catastrophic system failure, such as structural damage, aerodynamic stall, or loss of control \cite{Oudin2017,Stougie2024-ym}. Given the increasing complexity of modern aerospace platforms, robust and theoretically grounded methods for FEP are crucial to ensure operational safety while enabling vehicles to fully realize their performance potential. In most cases, FEP has been implemented at the attitude control level using reference filtering techniques \cite{Tang2009,Falkena2010}. Two widely used approaches in this context are reference clipping (RC) via saturation functions and model-based command filtering (MBCF). The first approach, RC, applies saturation functions that limit the reference commands based on predefined upper and lower bounds. This method ensures that commanded values do not exceed safe limits, effectively enforcing a constraint at the reference input level \cite{Seo2017}. While straightforward to implement, the RC approach fundamentally transforms the FEP problem into a reference tracking problem, leading to several limitations. Most notably, it does not account for the closed-loop system dynamics, which can result in constraint violations during transient phases, as observed in \cite{Stougie2024-ym, Steffensen2019}. Additionally, since the clipping function operates independently of the system dynamics, it lacks theoretical guarantees of strict constraint satisfaction under disturbances and unmodeled dynamics. An alternative method is MBCF, such as command filters or model reference filters \cite{Lombaerts2017}. These approaches incorporate an assumed dynamic model of the vehicle to shape the reference commands, enabling smoother control actions that respect dynamic constraints. When properly designed, such filters can provide empirical and theoretical guarantees of forward invariance, ensuring that the system remains within safe operating limits, as shown in \cite{Grondman2018}, \cite{Autenrieb_2024b} and \cite{Autenrieb2024}. However, the complexity of designing these filters increases significantly, particularly for systems with relative degrees greater than two.\\

Recent advancements in control theory offer an alternative perspective on designing FEP systems, framing the problem as a closed-loop state-constrained control problem. In such a framework, the controllers not only track desired set points but also dynamically enforce safety constraints throughout the system's operation, both during steady-state and transient phases. Control Barrier Functions (CBFs) have emerged as a powerful tool within this class of safety-critical control methods. CBFs guarantee the forward invariance of a defined safe set, providing formal assurances that the system remains within prescribed limits under all operational conditions \cite{Ames_2014}. The optimization-based nature of CBFs allows for seamless integration of a wide range of constraints, including physical limitations such as load factor and stall prevention, as well as operational constraints like altitude limits and obstacle avoidance \cite{Agrawal2021-aw, autenrieb2025_arxiv}. A significant advantage of CBF-based methods is their ability to augment existing control architectures. Rather than requiring a complete redesign, CBFs can function as an additional safety layer, leveraging the existing closed-loop dynamics as part of the optimization problem \cite{Ames_2017, Li2023-wy}. This compatibility facilitates their implementation in modern aerospace systems, while the real-time adaptability of CBFs enables optimal performance under varying conditions, disturbances, and uncertainties \cite{Taylor2020-em, Ames_2014, solanocastellanos2024}. Despite their potential, the application of CBFs to FEP remains relatively underexplored in the flight control community.\\

This paper presents a novel method for addressing the FEP problem by formulating it as a state-constrained optimization problem using CBFs. The proposed framework is designed to ensure system safety during transient and steady-state operations by explicitly considering the dynamics of the closed-loop system. Furthermore, the methodology highlights the ability of CBFs to dynamically enforce constraints, thereby offering a flexible solution that enhances the robustness of existing control architectures. To illustrate the efficacy of this approach, we derive a nonlinear flight controller for a missile vehicle and develop a CBF-based flight envelope protection system that dynamically enforces safety constraints. Numerical simulations validate the approach, comparing its performance against traditional methods such as reference signal clipping and saturation functions. These comparisons highlight the method’s superiority in addressing inherent limitations of classical approaches, including the absence of theoretical safety guarantees and overly conservative designs.\\

The remainder of this paper is organized as follows: Section~\ref{sec:Background} provides the theoretical foundation of CBFs and their key properties. Section~\ref{sec:he Flight Envelope Protection Problem} formulates the flight envelope protection problem, incorporating state and input constraints. In Section~\ref{sec:Nonlinear Flight Control System Design with Guranteed FEP}, a CBF-based control strategy is developed for a representative flight vehicle. Section~\ref{sec:Numerical Results} presents numerical simulations that evaluate the effectiveness of the proposed approach. Finally, Section~\ref{sec:Conclusion} summarizes the findings and discusses potential avenues for future research.

\section{Background}
\label{sec:Background}
To provide a clear understanding of the proposed and integrated approach, this section begins with a discussion of the mathematical concept of CBFs, which are fundamental to ensuring system safety through the enforcement of state constraints. Subsequently, we introduce the nonlinear missile dynamics considered in this work, providing the necessary details about the considered vehicle to establish the foundation for the later proposed FEP approach.

\subsection{The Mathematical Concept of Control Barrier Functions}

Consider a state space \( \chi \subset \mathbb{R}^n \), where it is assumed that \( \chi \) is path-connected and $ 0 \in \chi $. We first consider the following nonlinear system given by:
\begin{equation}
    \label{NonlinearPlant1}
    \dot{x}(t) = f(x(t)),
\end{equation}
where $x(t) \in \chi$ and $f: \chi \to \mathbb{R}^n$ is Lipschitz continuous. To define safety, we consider a continuously differentiable function  $h: \chi \rightarrow \mathbb{R}$ where $\chi \subset \mathbb{R}^n$, and a set $S$ defined as the zero-superlevel set of $h$, yielding:
\begin{align}
    S &\triangleq \begin{Bmatrix} x \in  \chi | h(x(t)) \geq 0 \end{Bmatrix}, \label{Safe_set_1}\\
   \partial S &\triangleq \begin{Bmatrix} x \in  \chi | h(x(t)) = 0 \end{Bmatrix}, \label{Safe_set_2}\\
   int(S) &\triangleq \begin{Bmatrix} x \in  \chi | h(x(t)) > 0 \end{Bmatrix}. \label{Safe_set_3}
\end{align}
We refer to $S$ as the safe set of the considered dynamics, given in Eq.~\eqref{NonlinearPlant1}. The following definitions are introduced \cite{Nagumo_1942, Blanchini_1999}:
\begin{definition}
The set $S$ is positively invariant for the system \eqref{NonlinearPlant1}, if for every $x_0 \in S$, it follows $x(t) \in S$ for $x(0) = x_0$ and all $t \in I(x_0) = [0,\infty)$.
\end{definition}
\begin{remark}
In the work presented here, the term forward invariant is also used. It is equivalent to positively invariant, and both can be used interchangeably in the context of CBFs. Ultimately, a positively invariant set under a CBF-based controller ensures that the system remains within the safe set indefinitely, guaranteeing adherence to predefined safety constraints.
\end{remark}
\begin{definition}
The set $S$ is weakly positively invariant for the system \eqref{NonlinearPlant1}, if among all the solutions of \eqref{NonlinearPlant1} originating in $x_0 \in S$, there exists at least one globally defined solution $x(t)$ which remains inside $S$ for all $t \in I(x_0) = [0,\tau_{max} = \infty)$.
\end{definition}
Next, we define the distance from a point to a set:
\begin{definition}
\label{distance_definition}
Given a set $S \subset \mathbb{R}^n$ and a point $y \subset \mathbb{R}^n$, the distance from the point to the set is defined as
\begin{equation}
    dist(y,S) = \inf_{w \in S} \| y-w \|_*
\end{equation}
where $\| \cdot \|_*$ is any relevant norm.
\end{definition}
Based on Definition \ref{distance_definition}, we can define a tangent cone for a closed set.
\begin{definition}
\label{tangent_cone_definition}
Given a closed set $S$, the tangent cone to $S$ at $x$ is defined as:
\begin{equation}
    T_S (x(t)) = 
    \begin{Bmatrix} 
    z: \liminf\limits_{\tau \tends 0} \frac{dist(x(t)+\tau z,S)}{\tau}=0
    \end{Bmatrix}.
\end{equation}
\end{definition}
If $S$ is convex $T_S(x(t))$ is convex, and “$\liminf$” can be replaced by “$\lim$”. Furthermore if $x \in int(S)$, then $T_S (x(t)) = \mathbb{R}^n$, whereas if $x \notin S$, then $T_S (x(t)) = \emptyset$, since $S$ is defined as a closed set. Therefore, $T_S (x(t))$ is only non-trivial on the boundary of $S$.

\begin{figure}
    \centering
    \includegraphics[width=0.6\linewidth]{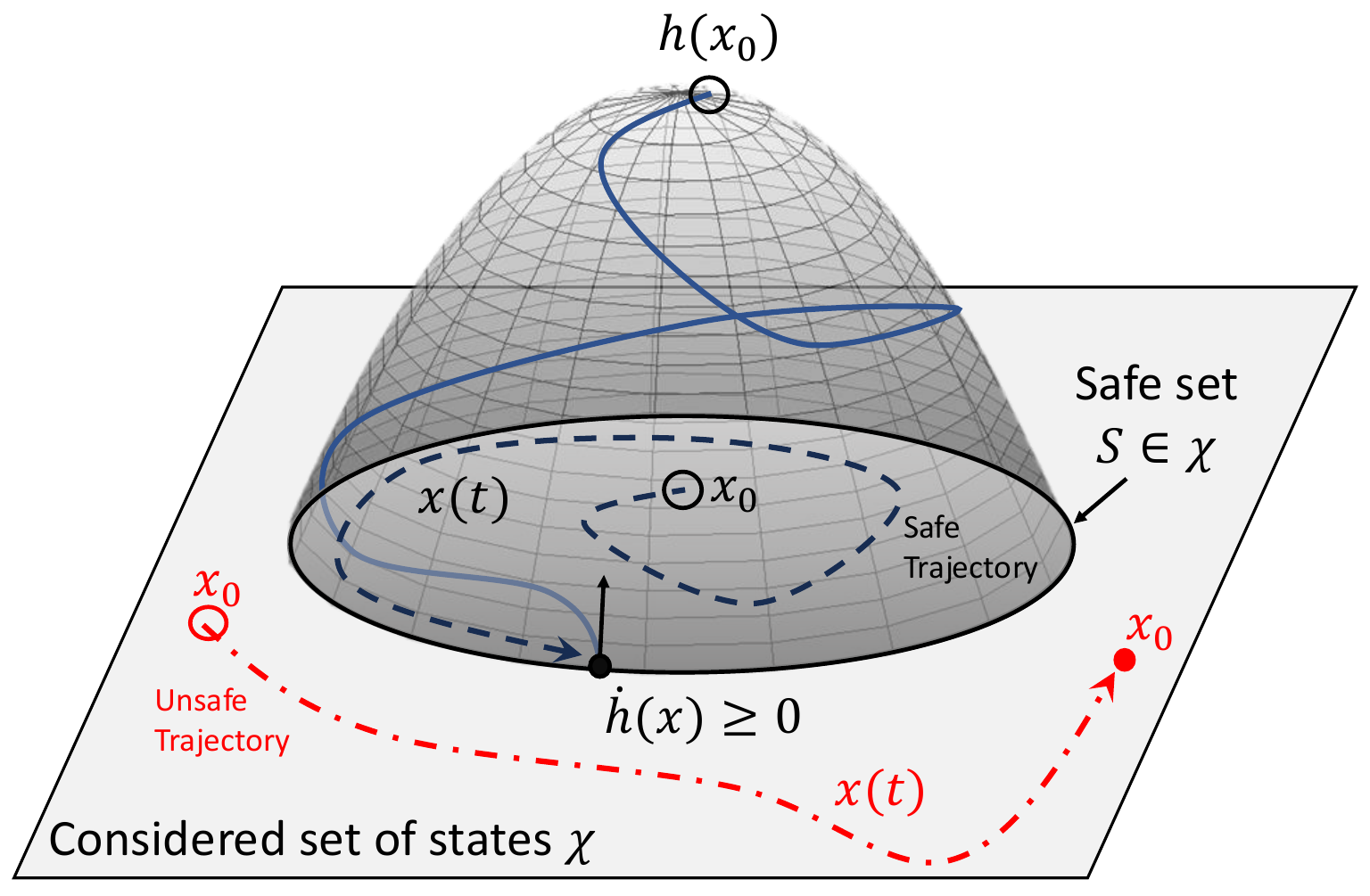}
    \caption{Illustration of two trajectories within $\chi$ but not within the safe set $S$ in dash-dotted red and one trajectory within the safe set $S$ in dashed black, together with its barrier function $h(x(t))$ in solid blue. }
    \label{fig:CBF_concept}
\end{figure}

We use Definition \ref{tangent_cone_definition} to introduce Nagumo's theorem\cite{Nagumo_1942}:
\begin{theorem}
\label{Nagumos_theorem_number}
Consider the system defined in \eqref{NonlinearPlant1}. Let $S \subset \mathbb{R}^n$ be a closed set. Then, $S$ is weakly positively invariant for the system if and only if \eqref{NonlinearPlant1} satisfies the following condition:
\begin{equation}
    \label{Nagumos_theorem}
    f(x(t)) \in T_S (x(t)), \,\,\, \text{for} \,\,\, \forall x(t) \in S.
\end{equation} 
\end{theorem}
The theorem states that if the direction of the dynamics defined in \eqref{NonlinearPlant1} for any $x(t)$ at the boundary of the safe set $\partial S$ points tangentially or inside the safe set $S$, then the trajectory $x(t)$ stays in $S$. 
\begin{definition}
A continuous function $\alpha: (-b, a) \rightarrow \mathbb{R}$, with $a,b > 0$, is an extended class $\mathcal{K}$ function $(\alpha \in \mathcal{K})$, if $\alpha(0) = 0$ and $\alpha$ is strictly increasing. If $a, b = \infty$, $\lim_{r \rightarrow \infty} \alpha(r) = \infty$, $\lim_{r \rightarrow -\infty} \alpha(r) = -\infty$ then $\alpha$ is said to be a class $\mathcal{K}_{\infty}$ function $(\alpha \in \mathcal{K}_{\infty})$.
\end{definition}
\begin{definition}
For the system considered in \eqref{NonlinearPlant1}, a continuously differentiable and convex function $h: \mathbb{R}^n \rightarrow \mathbb{R}$ is a zeroing barrier
function (ZBF) for the set $S$ defined by Eqs. \eqref{Safe_set_1}-\eqref{Safe_set_3}, if there exist an extended class $\mathcal{K}$ function $\alpha(h(x(t)))$ and a set $S \in \mathbb{R}^n$ such that $\forall x(t) \in S$,
\begin{equation}
    \label{ZeroingBarrierFunction}
    \dot{h}(x(t)) \geq -\alpha(h(x(t))).
\end{equation}
\end{definition}
The above definitions lead to a less restrictive version of Eq.~\eqref{Nagumos_theorem} as it weakens the requirement to that in Eq.~\eqref{ZeroingBarrierFunction}. 
The derived concept is illustrated in Fig.~\ref{fig:CBF_concept}. By definition, a valid barrier function $h(x(t))$ is defined as positive within the safe set $S$. To provide safety guarantees and forward invariance within $S$, the condition on the boundary of the safety set needs to be $\dot{h}(x(t))\geq 0$. When this condition is satisfied, it can be proven that the system remains safe. 
We now expand the scope of the problem statement from Eq.~\eqref{NonlinearPlant1} to those with an affine control input of the form:
\begin{equation}
    \label{NonlinearPlant2}
    \dot{x} = f(x(t)) + g(x(t)) u(t)
\end{equation}
where  $g: \chi \to \mathbb{R}^{n \times m}$ is Lipschitz and  $u(t) \in \mathbb{R}^m $. We introduce the notion of a CBF such that its existence allows the system to be rendered safe w.r.t. $S$ \cite{Ames_2014,Ames_2017} and allows a weaker requirement for system safety with a control input $u(t)$, similar to \eqref{ZeroingBarrierFunction}.
\begin{definition}
Let $S \subset \chi$ be the zero-superlevel set of a continuously differentiable function $h: \chi \rightarrow \mathbb{R}$. The function $h$ is a CBF for $S$ for all $x \in S$, if there exists a class $\mathcal{K}_{\infty}$ function $\alpha(h(x(t)))$ such that for the system defined in Eq.~\eqref{NonlinearPlant2} we obtain:
\begin{equation}
    \label{ControlBarrierFunction_simple}
    \sup_{u\in \mathbb{R}^m} \dot{h}(x(t),u(t))  \geq -\alpha(h(x(t))),
\end{equation}
where
\begin{equation}
\dot{h}(x(t),u(t)) = \frac{\partial h}{\partial x}\begin{bmatrix} f(x(t)) + g(x(t))u(t) \end{bmatrix} =L_{f} h(x(t)) + L_{g} h(x(t)) u(t)
\vspace{0.2cm}
\end{equation}
\end{definition}
As discussed in \cite{XU2015,Kolathaya2019}, any Lipschitz feedback controller in the form of:
\begin{equation}
    K_{S} (x(t)) = \{ u \in \mathbb{R}^m  : L_{f} h(x(t)) + L_{g} h(x(t)) u(t) + \alpha(h(x(t))) \geq 0 \}
\end{equation}
renders the system forward invariant within $S$. One way to obtain such a controller is to use the universal input-to-state safeguarding formula, as proposed in \cite{Kolathaya2019}, which delivers the following control law:
\begin{equation}  
k(x(t)) = \begin{dcases*}
     K_G(x(t))        & if $B(x(t)) \neq 0$,\\
    0  & if $B(x(t)) = 0$
\end{dcases*}
\end{equation}
where
\begin{equation}
\label{ContinousController}  
K_G(x(t)) = \frac{-A(x(t)) + \sqrt{A(x(t))^2 + |B(x(t))|^4}}{B(x(t))^T B(x(t))} B(x(t)) + B(x(t))
\end{equation}
\vspace{0.2cm}
with $A(x(t)) = L_{f} h(x(t)) + \alpha(h(x(t)))$  and $B(x(t)) = L_{g} h(x(t))^T$, which leads to $\dot{h}(x(t),u(t)) \geq 0$ at the boundaries. Hence, the system always stays within the safe set $S$. Alternatively, the authors in \cite{Ames_2014} proposed a real-time optimization-based safety filter to balance control performance and safety optimally. The formulated CBF-based quadratic program (QP-CBF) problem minimizes the Euclidean norm between a desired control input $u*(t)$ (for example, coming from a stable and performance-oriented nonlinear controller) and the safe control input $u(t)$:
\begin{argmini*}
    {u(t)\in \mathcal{\mathbb{R}}^m}{  	\lVert u(t) - u^*(t) 	\lVert_2^2}
    {}{}
    \addConstraint{L_{f} h(x(t)) + L_{g} h(x(t)) u(t) + \alpha(h(x(t))) \geq 0}
    \addConstraint{ A_u u(t) + b\leq 0},
\end{argmini*}
with $A_u u(t) + b$ representing box constraints on the control input $u$ to address the problem of a limited admissible set.

Since the relevant nonlinear dynamics of the flight systems considered in this work are of a relative degree greater than one, we extend the discussion of CBFs with the notion of High Order Barrier Functions (HOBFs) and High Order Control Barrier Functions (HOCBFs), as proposed in \cite{Xiao_2022hocbf}.  {\color{black} The control-theoretical construct of CBFs introduced
here, along with their extension to HOCBFs, provide a unified methodology for enforcing state constraints derived from operational limits, such as
aerodynamic, structural, and performance constraints, which are critical for ensuring safe flight with complex and nonlinear behaviors.} By establishing these theoretical foundations, the framework sets the stage for its application to practical systems, as demonstrated in this work through its integration into an FEP scheme. we consider a time-varying function to define an invariant set for system \eqref{NonlinearPlant1}. For a $d$-th order differentiable function $h : \mathbb{R}^n \times [t_0, \infty) \to \mathbb{R}$ (where $t_0$ denotes the initial time), we define a series of functions $\psi_0, \psi_1, \ldots, \psi_d$ recursively as:
\begin{align}
    \psi_0(x(t), t) &= h(x(t), t), \notag \\
    \psi_1(x(t), t) &= \dot{\psi}_0(x(t), t) + \alpha_1(\psi_0(x(t), t)),\label{eqn:HOCBF_safe_sets} \\
    \psi_2(x(t), t) &= \dot{\psi}_1(x(t), t) + \alpha_2(\psi_1(x(t), t)), \notag\\
    &\vdots \nonumber \\
    \psi_d(x(t), t) &= \dot{\psi}_{d-1}(x(t), t) + \alpha_d(\psi_{d-1}(x(t), t)), \notag
\end{align}
where $\alpha_1, \alpha_2, \ldots, \alpha_d$ are class $\mathcal{K}$ functions. We further define a series of sets $S_1(t), S_2(t), \ldots, S_d(t)$ associated with \eqref{eqn:HOCBF_safe_sets} in the form:
\begin{align}
    S_1(t) &= \{x \in \mathbb{R}^n : \psi_0(x(t), t) \geq 0\}, \\
    S_2(t) &= \{x \in \mathbb{R}^n : \psi_1(x(t), t) \geq 0\}, \\
    &\vdots \nonumber \\
    S_d(t) &= \{x \in \mathbb{R}^n : \psi_{d-1}(x(t), t) \geq 0\}.
\end{align}
The forward invariance of these sets collectively ensures that the trajectory of system \eqref{NonlinearPlant1} remains within a safe region for all $t \geq t_0$.
\begin{definition}
\label{higher_order_safety_requirement}
A function $h(x(t), t)$ is a High Order Barrier Function (HOBF) of relative degree $d$ for \eqref{NonlinearPlant1} if it satisfies:
\begin{equation}
\label{higher_order_safety_requirement_equation}
    \psi_d(x(t), t) \geq 0,
\end{equation}
for all $(x(t), t) \in S_1(t) \cap S_2(t) \cap \cdots \cap S_d(t) \times [t_0, \infty)$. This guarantees the forward invariance of \eqref{NonlinearPlant1} in the set $S_1(t) \cap S_2(t) \cap \cdots \cap S_d(t)$.
\end{definition}

When a control-affine system of the form \eqref{NonlinearPlant2} is considered, HOBFs can be extended to High Order Control Barrier Functions (HOCBFs) by incorporating control inputs.
\begin{definition}
\label{higher_order_safety_requirement2}
Let $S_1(t), S_2(t), \ldots, S_d(t)$ be defined by (12) and $\psi_0(x(t), t), \psi_1(x(t), t), \ldots, \psi_d(x(t), t)$ be defined by (11).
A function $h : \mathbb{R}^n \times [t_0, \infty) \to \mathbb{R}$ is a High Order Control Barrier Function (HOCBF) of relative degree $d$ for the system if there exist differentiable class $\mathcal{K}$ functions $\alpha_1, \alpha_2, \ldots, \alpha_d$ such that:
\begin{equation}
    L_f^d h(x(t), t) + L_g^{d-1} h(x(t), t) u + \frac{\partial^d h(x(t), t)}{\partial t^d} + O(h(x(t), t)) + \alpha_d(\psi_{d-1}(x(t), t)) \geq 0,
\end{equation}
for all $(x(t), t) \in S_1(t) \cap S_2(t) \cap \cdots \cap S_d(t) \times [t_0, \infty)$. In the above equation, $O(h(x(t), t))$ denotes the remaining Lie derivatives along $f$ and partial derivatives with respect to $t$ with degree less than or equal to $d - 1$.
\end{definition}
This framework introduces a systematic approach to ensure system safety by directly incorporating control inputs into the analysis. Unlike conventional reference filtering methods, HOCBFs enforce safety constraints as an inherent property of the system dynamics, ensuring that operational limits—such as those on angle of attack, load factor, or aerodynamic stability—are satisfied at all times. This makes HOCBFs particularly well-suited for applications in aerospace systems, where higher-order dynamics must be carefully managed to guarantee safe and efficient flight operations.

\subsection{Considered Nonlinear Missile Dynamics}
\label{Sect: Nonlinear_Missile_Model}
\begin{figure}[ht]
    \centering
    \includegraphics[width=0.6\textwidth]{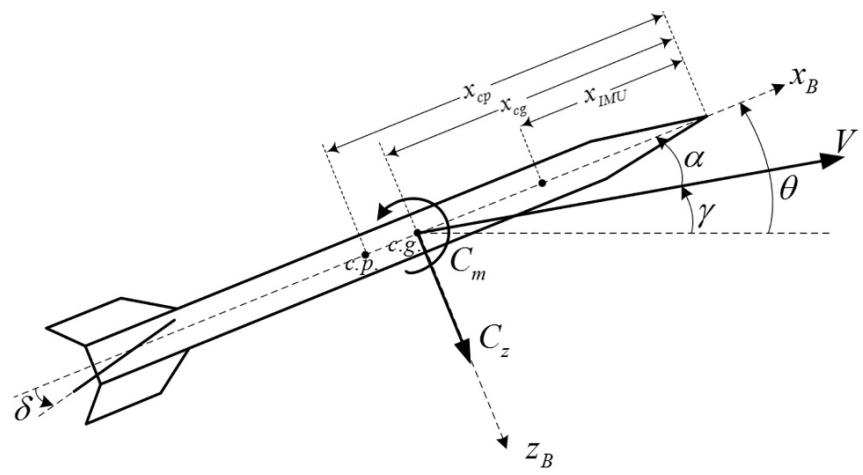}
    \caption{Sketch of external forces and moments acting on an axially symmetric missile.}
    \label{fig:missile_model}
\end{figure}

This study considers an axially symmetric skid-to-turn tail-controlled missile system, as illustrated in Fig.~\ref{fig:missile_model} ~ \cite{Lee_2016}. 

Under the assumption that the roll motion is rapidly stabilized, the missile motions can be decoupled into two perpendicular and identical channels (i.e., pitch and yaw). The missile pitch dynamics model utilized in this study is an adjusted version of the dynamical model shared in \cite{Hwang_2017}. The model is a reduced-order version of a six-degree-of-freedom model, where non-relevant dynamics have been removed. 

For the here presented work, the pitch motion is considered as follows:
\begin{align}
    \dot{\alpha}(t) &= \frac{QS}{mV} \left( C_{Z,\alpha}(M)\alpha(t)+C_{Z,\delta}(M)\delta (t)\right) + q(t), \label{eqn:alpha_dot} \\
    \dot{q}(t) &= \frac{QSd}{I_{yy}} \left(C_{M,\alpha}(M)\alpha(t)+C_{M,q}(M)\frac{d}{2V}q(t)+C_{M,\delta}(M)\delta(t) \right), \label{eqn:q_dot} \\
    n_z(t)&= \frac{QS}{mg} \left(C_{Z,\alpha}(M)\alpha(t)+C_{Z,\delta}(M)\delta(t)\right), \label{eqn:az}
\end{align}
where \(\alpha(t)\) and \(q(t)\) represent the angle of attack and the pitch rate, which are the state variables. Furthermore, \(\delta(t)\) represents the control fin deflection, which is the control input of the missile system, and \(n_z\) denotes the lateral load. V represent the missile velocity \(Q\) represents the dynamic pressure (\(\rho V^2 / 2\)). The aerodynamic coefficients \(C_{Z,\alpha}\), \(C_{Z,\delta}\),\(C_{M,\alpha}\), \(C_{M,q}\), and \(C_{M,\delta}\) are obtained from wind-tunnel testing or aero-prediction methods. \(S\), \(d\), \(m\), and \(I_{yy}\) refer to the constant reference area, diameter, mass, and moment of inertia, respectively.

{\color{black}
The integrated actuator dynamics are modeled as first-order systems with predefined control surface deflection and rate limits. The relevant environmental, aerodynamic coefficients, and vehicle parameters used for the flight dynamical modeling in this study are provided in Table~\ref{tab:missile_params}.
\begin{table}[hbt!]
\caption{\label{tab:missile_params} Flight vehicle parameters, aerodynamic coefficients, and constraints.}
\centering
\begin{tabular}{lcc}
\hline
Parameter & Symbol & Value \\
\hline
Velocity & $V$ & 914.00 m/s \\
Air density & $\rho$ & 1.225 kg/m$^3$ \\
Mass & $m$ & 453.00 kg \\
Moment of inertia & $I_{yy}$ & 1407.00 kgm$^2$ \\
Gravity & $g$ & 9.80665 m/s$^2$ \\
Mach number & $M$ & 2.6859 \\
Reference area & $S$ & 0.073 m$^2$ \\
Missile diameter  & $d$ & 0.30 m \\
\hline
Aerodynamic Coefficients & & \\
\hline
$C_{Z,\alpha}$ & & -32.5925 \\
$C_{Z,\delta}$ & & -7.1863 \\
$C_{m,0}$ & & -3.277 \\
$C_{m,\alpha}$ & & -80.4716 \\
$C_{m,q}$ & & -56.1499 \\
$C_{m,\delta}$ & & -69.6272 \\
\hline
Control Input Constraints & & \\
\hline
Fin deflection limits & $\delta$ & $\pm 30^\circ$ \\
Fin deflection rate limits & $\dot{\delta}$ & $\pm 90^\circ$/s \\
\hline
\end{tabular}
\end{table}
}

\section{The Flight Envelope Protection Problem}
\label{sec:he Flight Envelope Protection Problem}
\begin{figure}[ht]
    \centering
    \includegraphics[width=0.9\columnwidth]{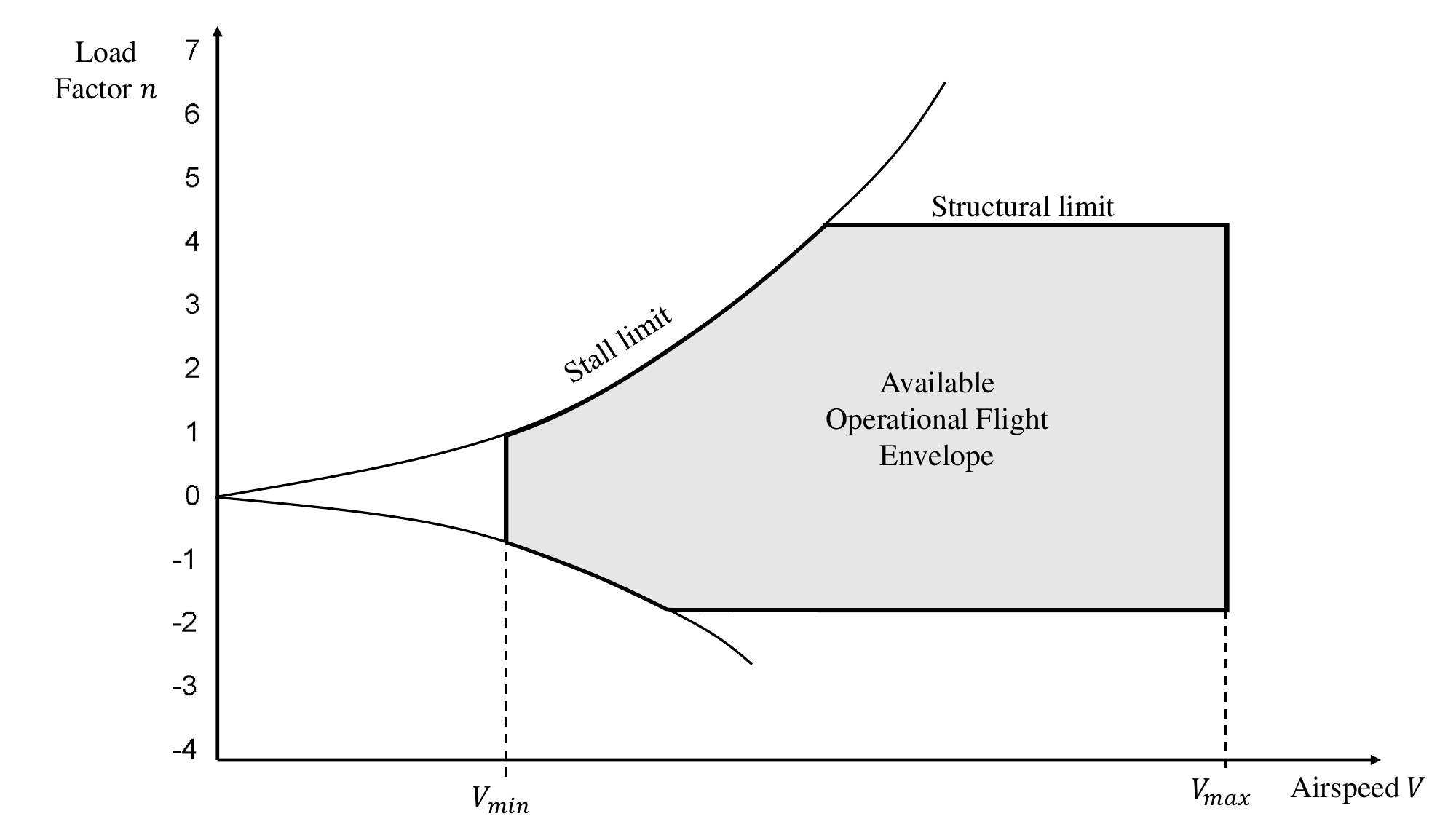}
    \caption{Example of a V-n diagram illustrating the relationship between load factors and airspeed, with limits defined by stall and structural loads.}
    \label{fig:V-n_Diagram}
\end{figure}

For flight systems, aerodynamics are the primary driver of the vehicle's dynamics, governing its performance and stability during flight. Aerodynamic forces and moments arise from the interaction between the vehicle and the surrounding air, affected by control surfaces and incidence angles such as the angle of attack \(\alpha(t)\) and sideslip angle \(\beta(t)\). A flight envelope, for any flight vehicle, can be effectively defined and graphically illustrated using a V-n diagram, as shown in Fig.~\ref{fig:V-n_Diagram}. This diagram illustrates the upper and lower load limits corresponding to different airspeeds. These limits provide critical insights into the vehicle's operational constraints and form the basis for understanding load restrictions.

From the V-n diagram, it is evident that the load limits \(n_{z, \text{min}}(V)\) and \(n_{z, \text{max}}(V)\) are functions of the airspeed \(V\). The operational airspeed \(V\) is bounded by \(V_{\text{min}}\) and \(V_{\text{max}}\). Classically, \(V_{\text{min}}\) represents the lowest airspeed at which stall can be avoided at the maximum allowable angle of attack \(\alpha\). Conversely, \(V_{\text{max}}\) represents the maximum airspeed, typically limited by the structural integrity of the vehicle, for example, to prevent flutter effects in conventional aircraft configurations. These bounds ensure the vehicle operates safely within a defined range. The load limits \(n_{z, \text{min}}(V)\) and \(n_{z, \text{max}}(V)\) are used to define a speed-dependent safe set for the load factor:

\begin{equation}
S(V) \;=\; \bigl\{\,n_z\in\mathbb R \;\big|\;
    n_{z,\min}(V)\le n_z \le n_{z,\max}(V)
\bigr\},
\quad V\in [V_{\min},\,V_{\max}].
\end{equation}

Since \(V\) is treated as a separate control problem and varies more slowly than the loads, we can, without loss of generality, assume a pointwise quasi-static load limit at each time step \(t\). This means that for each time \(t\), the load limits \(n_{z, \text{min}}\) and \(n_{z, \text{max}}\) can be considered constant during the controller's evaluation, ensuring computational feasibility. Under this assumption, the safety set becomes time-dependent:
\begin{equation}
    S(t) := \{n_z(t) \in \mathbb{R} \mid n_{z, \text{min}}(t) \leq n_z(t) \leq n_{z, \text{max}}(t) \}, \quad \forall t \geq t_0.
\end{equation}

Consequently, to ensure safety with respect to the defined flight envelope, the system must satisfy the following forward invariance property:
\begin{equation}
    x(t_0) \in S(t) \implies x(t) \in S(t), \quad \forall t \geq t_0,
\end{equation}
where \(S(t)\) is the safe set defined by the time-dependent load constraints. 

In the following sections, we will demonstrate how the presented FEP problem can be generalized as a state-constrained control problem. Furthermore, we will illustrate how CBFs provide an effective framework for achieving this goal.

\section{Nonlinear Flight Control System Design with Guaranteed FEP}
\label{sec:Nonlinear Flight Control System Design with Guranteed FEP}
{\color{black}
This section presents the design of a nonlinear flight control system integrated with a FEP safety filter. The overall architecture of the established control system is illustrated in a block diagram in Fig.~\ref{fig:CBF_Filter_Block_Diagram}.

\begin{figure}[ht]
    \centering
    \includegraphics[width=1\linewidth]{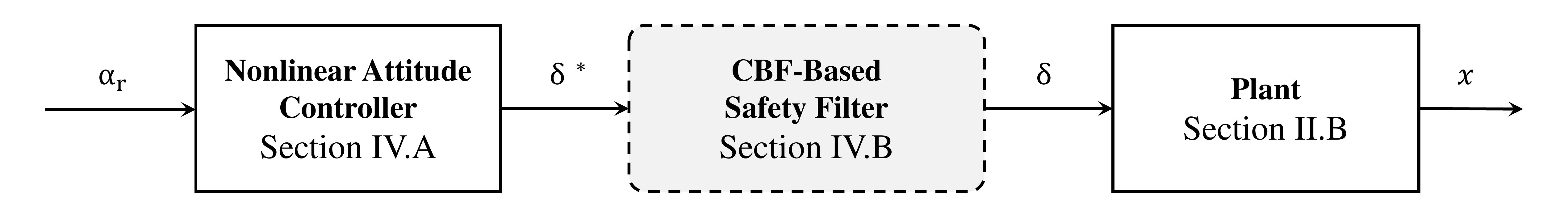}
    \caption{Block diagram of the integrated control architecture with a FEP safety filter.}
    \label{fig:CBF_Filter_Block_Diagram}
\end{figure}

First, a cascaded nonlinear attitude flight controller is designed in Sect.~\ref{sec:Nonlinear Attitude Controller Design} to regulate the nonlinear longitudinal missile dynamics. Following this, a CBF-based safety filter is proposed in Sect.~\ref{Sect: CBF-Based Nonlinear Safety Filter Design} to ensure adherence to predefined safety constraints, effectively maintaining the missile system's operation, defined in Sect~\ref{Sect: Nonlinear_Missile_Model}, within the defined flight envelope while minimally influencing control performance when the dynamics are far from the envelope boundaries.
}
\subsection{Nonlinear Attitude Controller Design}
\label{sec:Nonlinear Attitude Controller Design}
For the sake of reproducibility of the results presented here, it was decided to control the nonlinear longitudinal dynamics of the missile using a simple cascaded nonlinear dynamic inversion (NDI) control strategy, as presented in \cite{daCosta2003}. However, it is worth noting that the choice of attitude controller can be viewed as separate from the FEP problem, and any performance-oriented attitude controller could be integrated into the proposed overall flight control architecture. 

\begin{figure}[ht]
    \centering
    \includegraphics[width=0.9\textwidth]{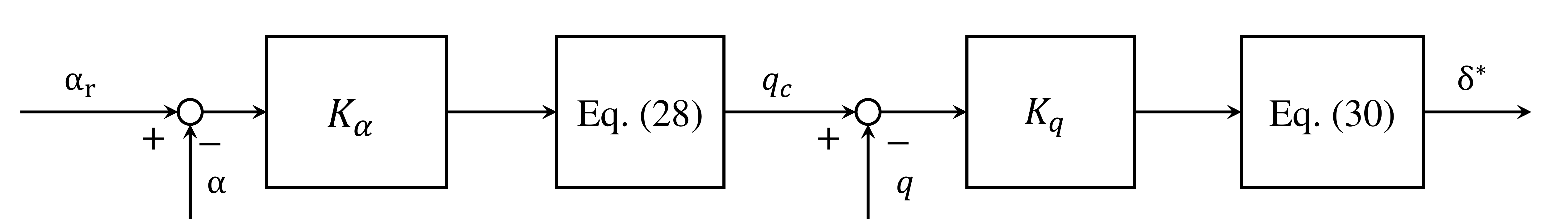}
    \caption{Cascaded inversion-based controller for longitudinal missile dynamics.}
    \label{fig:inversion_controller}
\end{figure}

The fundamental principle of NDI-based control systems is to transform the nonlinear input-output relationship into a linear relationship through state feedback and coordinate transformation. This method contrasts with the traditional Jacobian linearization, which relies on local linear approximations at specific operating points within the flight envelope \cite{Acquatella_2022}. By leveraging nonlinear control techniques, the requirement for gain scheduling is eliminated, resulting in a simplified system response with controlled variables exhibiting integrator-like dynamics \cite{Holzapfel_2004}. The cascaded controller architecture is depicted in Fig.~\ref{fig:inversion_controller}.
The control design begins by regulating the angle of attack dynamics via:
\begin{equation}
    \dot{\alpha}_c (t)  = K_{\alpha} (\alpha_r(t) - \alpha(t)),
    \label{eqn:dot_alpha_c}
\end{equation}
where \(\alpha_r(t)\) is the reference angle of attack that is aimed to be tracked, \(\alpha(t)\) is the measured angle of attack, and \(K_{\alpha}\) represents a positive feedback gain. Given the relationship in Eq.~\eqref{eqn:q_dot}, Eq.~\eqref{eqn:dot_alpha_c} can be transformed into a pitch rate command \(q_c(t)\)  for the inner-loop controller with the following inversion law:
\begin{equation}
    q_c(t) = \dot{\alpha}_c(t) - \frac{QS}{mV} \left(C_{Z,\alpha}(M) \, \alpha(t) +C_{Z,\delta}(M ) \, \delta(t) \right).
\end{equation}
The linear inner-loop controller calculates the pitch rate time derivative $\dot{q}_c(t)$ needed to track the obtained pitch rate reference signal $q_c(t)$ from the prior angular inversion step using:
\begin{equation}
    \dot{q}_c(t)  = K_{q} (q_c(t) - q(t)),
\end{equation}
where \(q(t)\) is the measured pitch rate output and \(K_q\) denotes the corresponding positive feedback gain. Finally, the performance-oriented, but potentially unsafe, control surface deflection \(\delta^*\) is determined by using the nonlinear dynamics of Eq.~\eqref{eqn:q_dot} as:
\begin{equation}
    \delta^*(t) = C_{M,\delta}(M )^{-1} \left(\dot{q}_c(t)  - \frac{QSd}{I_{yy}} \left(C_{M,\alpha}(M)\alpha(t)  + C_{M,q}(M)\frac{d}{2V} q(t) \right)\right).
\end{equation}

\begin{figure}[ht]
    \centering
    \includegraphics[width=1\linewidth]{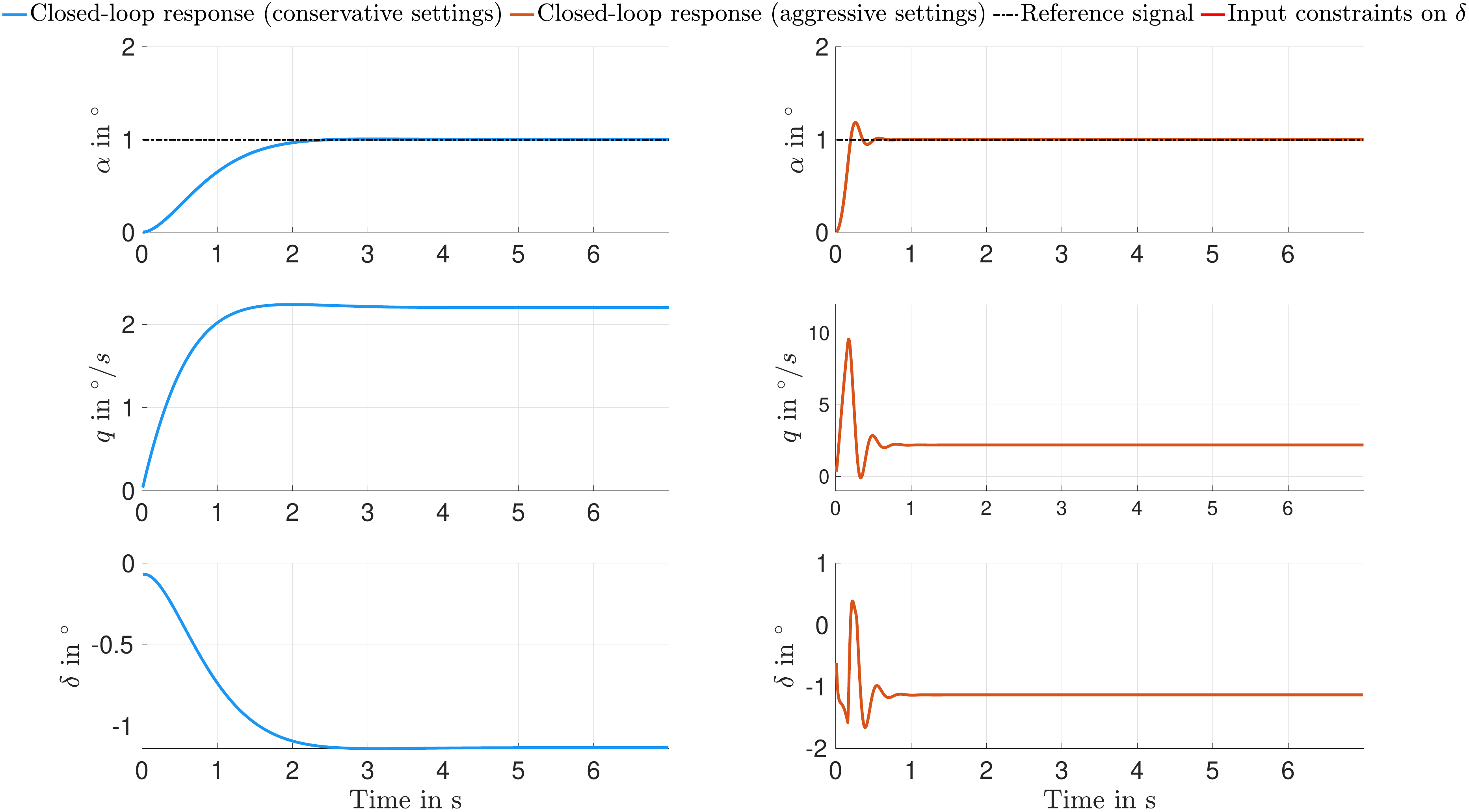}
    \caption{Closed-loop response of the cascaded NDI controller to a reference command of \( \alpha_r(t) = 1^\circ \). Left: conservative control setting with steady-state error and slow tracking. Right: aggressive control setting with fast response and moderate overshoot.}
    \label{fig:NDI_Comparison}
\end{figure}

As discussed in Section \ref{sec:Introduction}, two widely used methods in classical FEP are RC via saturation functions and MBCF. While these two methods aim to enforce constraints by shaping the commanded trajectories, their effectiveness depends critically on the dynamic behavior of the closed-loop system. In particular, tracking delays, steady-state errors, or overshoots in response to reference commands can result in constraint violations—even when the reference signal itself remains within nominally safe bounds. To explicitly analyze this dependency later on in this work, two distinct gain configurations of the previously introduced cascaded NDI controller are considered. The controller gains have been defined using a standard Linear Quadratic Regulator (LQR) technique \cite{Kalman_1960LQR}. The first gain configuration is referred to as the "conservative" control setting, characterized by lower gain values. While this tuning results in a conservative and less aggressive control response, it leads to slow reference tracking and smaller overshoots in tracking. The second configuration is labeled the "aggressive" control setting, which uses higher gain values to yield faster tracking performance. In this case, the system response is considerably more agile, closely following the reference signal with minimal phase lags, albeit at the cost of a transient overshoot. The relevant parameters for the considered controller settings are given in Table~\ref{tab:controll_gains}.
\begin{table}[hbt!]
\caption{\label{tab:controll_gains} Selected gain parameters for the cascaded NDI controller under conservative  and aggressive tuning configurations.}
\centering
\begin{tabular}{lcc}
\hline
Controller Gain & Conservative & Aggressive \\
\hline
$K_\alpha$  & 4.00 & 36.53 \\
$K_q$       & 1.2 & 12.22 \\
\hline
\end{tabular}
\end{table}
The response behavior of both configurations is shown in Fig.~\ref{fig:NDI_Comparison}. The figure compares the time histories of angle of attack \(\alpha(t)\), pitch rate \(q(t)\), and control input \(\delta(t)\) for a reference command of \( \alpha_r(t) = 1^\circ \). On the left-hand side, the conservative controller exhibits a more sluggish response and limited tracking capability. In contrast, on the right, the aggressive controller provides rapid convergence to the reference at the expense of moderate overshoot and more aggressive control effort. It will later be demonstrated how this difference in controller performance results in inherent coupling between the required FEP guarantees and the controller’s tracking dynamics, and that this coupling has significant implications for the robustness and safety guarantees of conventional FEP schemes.

\begin{remark}
It should be noted that, as will be shown in the following section, the potentially unsafe control surface deflection \( \delta^*(t) \) will be adjusted to ensure that the flight vehicle remains within its designated flight envelope. However, this adjustment comes at the cost of deviating from the reference command \( \alpha_r(t) \). This deviation can introduce challenges, particularly when integral controllers are employed within the integrated flight control system to ensure adequate tracking performance. In such cases, integrator windup may occur, leading to undesirable control behavior and degraded system performance. To mitigate these effects, an {\color{black} windup protection approach} must be implemented to prevent excessive error accumulation and maintain stable control performance.
\end{remark}

\subsection{CBF-Based Nonlinear Safety Filter Design}
\label{Sect: CBF-Based Nonlinear Safety Filter Design}
In Sect.~\ref{sec:he Flight Envelope Protection Problem}, the flight envelope and its associated safety constraints were introduced based on airspeed-dependent load factor limits in the V-n diagram, cf. Fig.~\ref{fig:V-n_Diagram}. However, these constraints can also be expressed equivalently in terms of airspeed-dependent limits on the angle of attack \(\alpha(t)\). This transformation provides a more convenient representation for deriving control-theoretic safety constraints, as \(\alpha(t)\) is a directly regulated state variable in the integrated missile’s attitude control system, as presented earlier in this section.
\begin{figure}[ht]
    \centering
    \includegraphics[width=0.7\linewidth]{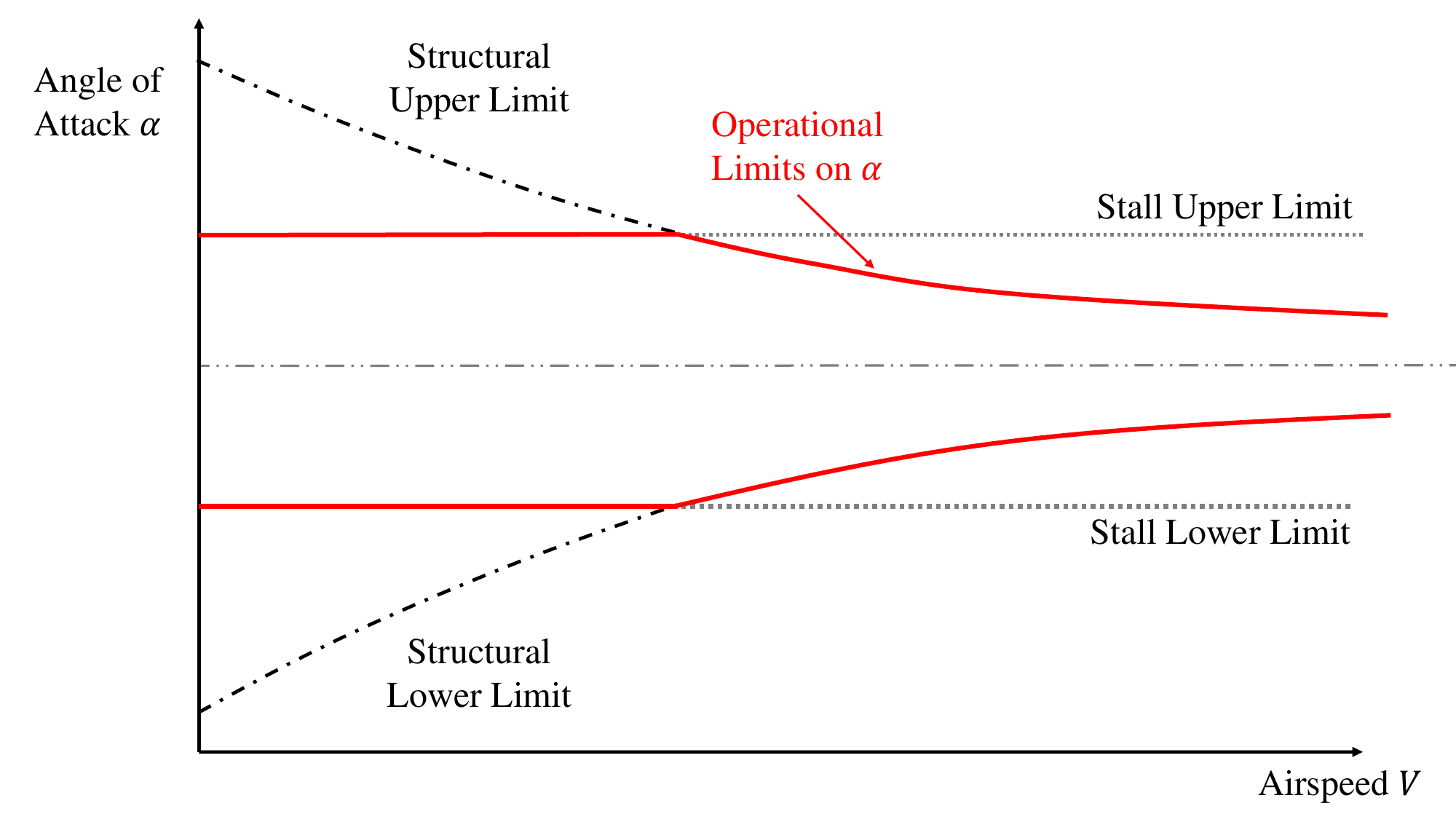}
    \caption{Qualitative relationship between the angle of attack $\alpha$ and airspeed $V$.}
    \label{fig:V-alpha_Diagram}
\end{figure}
Fig.~\ref{fig:V-alpha_Diagram} illustrates the qualitative relationship between airspeed \(V\) and the permissible limits on the angle of attack \(\alpha(t)\), which must not be violated to ensure safe operation. The dashed-dotted black lines represent structural constraints, the dotted black lines indicate stall boundaries, and the solid red lines delineate the combined overall operational limits of \(\alpha(t)\) as a function of airspeed. The operational limits on \(\alpha(t)\) imposed by structural load constraints can be derived from Eq.~\eqref{eqn:az}, considering the strictly defined structural load factor limits \( n_{z,\text{max,structural}} \) and \( n_{z,\text{min,structural}} \) the equation can be inverted to obtain the corresponding maximum and minimum permissible angles of attack \( \alpha_{\text{max,structural}} \) and \( \alpha_{\text{min,structural}} \). The structural constraints thereby impose an inverse-square dependency of \(\alpha(t)\) on \(V\), owing to the quadratic dependence of \(Q\) on \(V\). In contrast, stall limits are typically assumed to be independent of airspeed, meaning that the maximum and minimum allowable angles of attack due to stall, \( \alpha_{\text{max,stall}} \) and \( \alpha_{\text{min,stall}} \), remain constant across the flight envelope. As a result, the operational bounds on \(\alpha(t)\) are shaped by the interplay between structural and stall constraints. At lower airspeeds, stall limits dominate the constraints on \(\alpha(t)\). In contrast, at higher airspeeds, the range of feasible \(\alpha(t)\) values is increasingly restricted by structural considerations, as illustrated in Fig.~\ref{fig:V-alpha_Diagram} with solid red lines. By considering the described two origins of  constraints on $\alpha(t)$, the overall operational limits can be formally expressed as:
\begin{equation} 
    \alpha_{\text{crit,min}}(t) \leq \alpha(t)  \leq \alpha_{\text{crit,max}}(t) .
    \label{eqn:alpha_crit_constraints}
\end{equation}
where:
\begin{align}
    \alpha_{\text{crit,max}}(t)  &= \min(\alpha_{\text{max,stall}}(t) , \alpha_{\text{max,structural}}(t) ), \nonumber \\
    \alpha_{\text{crit,min}}(t) &= \max(\alpha_{\text{min,stall}}(t) , \alpha_{\text{min,structural}}(t)). \nonumber 
\end{align}
This property enables a piecewise simplification of the optimization problem as different constraints become active in distinct regions of the flight envelope. As outlined in Sect.~\ref{sec:he Flight Envelope Protection Problem}, the airspeed \(V\) can be assumed to vary relatively slowly. Consequently, the critical limits \( \alpha_{\text{crit,min}}(t) \) and \( \alpha_{\text{crit,max}}(t) \) also vary slowly. This justifies a pointwise quasi-static treatment, where their time dependence can be neglected for purposes of constraint formulation and control design. Allowing to formulate the safety problem such that $\alpha_r(t)$ remains within two pointwise static bounds $\alpha_{\text{crit,min}}$ and $\alpha_{\text{crit,max}}$. 

To dynamically enforce the derived angle-of-attack constraints, a HOCBF-based safety filter is implemented. The integrated filter modifies the performance-oriented and potentially unsafe control input \( \delta^*(t) \)  to ensure that the actual angle of attack \( \alpha(t)\) remains within the safe operational bounds \( \alpha_{\text{crit,min}}(t) \) and \( \alpha_{\text{crit,max}}(t) \) at all times, preventing violations of stall and structural load constraints. The safety filter optimally adjusts the control input \( \delta^*(t) \) by solving:  
\begin{align}
   &\min_{\delta, \dot\delta}
   \begin{aligned}[t]
      &\lVert \delta(t) - \delta^*(t) \rVert_2^2
   \end{aligned} \notag\\
   &\text{s.t.} \notag \\
   & {\psi}_{1,2}(\alpha(t))  \geq 0,  \notag\\ 
   & {\psi}_{2,2}(\alpha(t)) \geq 0, \label{eqn:QP_CBF_FEP}\\ 
   & \delta_{min}\leq \delta(t) \leq \delta_{max}, \notag\\
   & \dot\delta_{min}\leq \dot\delta(t) \leq \dot\delta_{max}. \notag  
\end{align}

Here, \( \delta^*(t) \) is the desired but potentially unsafe control input, and \( \delta(t) \) is the optimized safe input signal passed to the plant. The objective function ensures that \( \alpha_r(t) \) remains as close as possible to \( \alpha_r^*(t) \), minimizing unnecessary modification while enforcing safety. The HOCBF-derived safety constraints \( {\psi}_{1,2}(\alpha(t))  \geq 0\) and \( {\psi}_{1,2}(\alpha(t))  \geq 0 \) enforce the upper and lower angle-of-attack limits, ensuring \( \alpha(t) \) remains within \( \alpha_{\text{crit,min}} \) and \( \alpha_{\text{crit,max}} \). The final two inequality constraints make sure that \( \delta(t) \) stays within the admissible set of control inputs with respect to magnitude and rate limits.

\begin{remark}
In the integrated QP-based safety filter formulation, the upper and lower bounds on $\alpha(t)$ are enforced via two distinct HOCBF-derived inequality constraints. In general, this setup could lead to feasibility issues in the QP, especially when both constraints become simultaneously active or conflicting. However, due to the structure of the considered FEP problem, this conflict is typically avoided in practice. Specifically, the system either approaches the upper or the lower bound of the operational $\alpha(t)$ limits depending on the direction of motion—i.e., whether $\alpha(t)$ is increasing or decreasing. As a result, only one of the two HOCBF constraints is usually active at a given time, which significantly reduces the risk of mutual constraint conflict. Nevertheless, additional care must be taken to ensure feasibility when input constraints are present. As discussed in~\cite{Spiller_2025}, feasibility guarantees can be ensured through careful selection of the design parameters for the integrated CBFs. This is further discussed in the following part of this section, which outlines a tuning strategy to maintain forward invariance while preserving the feasibility of the integrated safety filter.
\end{remark}

\subsubsection{Derivation of CBF-based Safety Constraints}

While the QP framework has already been introduced in Eq.~\eqref{eqn:QP_CBF_FEP}, the explicit formulation of the constraints has not yet been discussed in detail. To address this, we construct HOCBF constraints that enforce the operational bounds on \(\alpha(t)\) dynamically, ensuring that the system remains within the permissible flight envelope at all times, as defined in Eq.~\eqref{eqn:alpha_crit_constraints}. To do so, we introduce the candidate barrier functions for the upper and lower angle-of-attack constraints:

\vspace{0.3cm}
\noindent\begin{minipage}{0.49\linewidth}
\begin{equation}
    \label{eqn:h_1}
     h_1(\alpha(t)) = \alpha_{\text{crit,max}}(t)  - \alpha(t) ,
\end{equation}
\end{minipage}
\hfill
\begin{minipage}{0.49\linewidth}
    \begin{equation}
    \label{eqn:h_2}
    h_2(\alpha(t)) = \alpha(t)  - \alpha_{\text{crit,min}}(t) .
    \end{equation}
\end{minipage}
\vspace{0.3cm}

These functions define the safe regions:

\vspace{0.3cm}
\noindent\begin{minipage}{0.49\linewidth}
\begin{equation}
    \label{eqn:S_1}
      S_1(t) = \{ \alpha(t)  \in \mathbb{R} : h_1(\alpha(t) ) \geq 0 \},
\end{equation}
\end{minipage}
\hfill
\begin{minipage}{0.49\linewidth}
    \begin{equation}
    \label{eqn:S_2}
    S_2(t) = \{ \alpha(t)  \in \mathbb{R} : h_2(\alpha(t) ) \geq 0 \}.
    \end{equation}
\end{minipage}
\vspace{0.3cm}

We begin by deriving the constraint for the upper bound on \(\alpha(t)\) and subsequently extend the formulation to the lower bound in an analogous manner based on the defined candidate from Eq.\eqref{eqn:h_1} for the associated safety set of Eq.~\eqref{eqn:h_2}.  As outlined before, it can be assumed \( \alpha_{\text{crit,min}}(t) \) and \( \alpha_{\text{crit,max}}(t) \) vary slowly, suggesting $\dot\alpha_{\text{crit,max}}(t) \approx 0$ and $\dot\alpha_{\text{crit,min}}(t)\approx 0$. Since \(\alpha(t)\) exhibits second-order dynamics, we construct a second-order HOCBF by recursively defining:
\begin{align}
    \psi_{1,0}(\alpha(t)) &= h_1(\alpha(t)), \\
    \psi_{1,1}(\alpha(t)) &= \dot{\psi}_{1,0}(\alpha(t)) + \gamma_{1,1} \psi_{1,0}(\alpha(t))  \,\,= -\dot{\alpha}(t) + \gamma_{1,1} \psi_{1,0}(\alpha(t)), \\
    \psi_{1,2}(\alpha(t))  &= \dot{\psi}_{1,1}(\alpha(t)) + \gamma_{1,2} \psi_{1,1}(\alpha(t)) 
      =  -\ddot{\alpha}(t) -\gamma_{1,1}\dot{\alpha}(t) + \gamma_{1,2} \psi_{1,1}(\alpha(t)) \label{eqn:final_HOCB_upper} \\& \,\,\,\,\,\,\,\,\,\,\,\,\,\,\,\,\,\,\,\,\,\,\,\,\,\,\,\,\,\,\,\,\,\,\,\,\,\,\,\,\,\,\,\,\,\,\,\,\,\,\,\,\,\,\,\,\,\,\,\,\,\,\,\,\,\,\,\,\,\,\,\,\, =  -\ddot{\alpha}(t) - (\gamma_{1,1}+  \gamma_{1,2})\dot{\alpha}(t) + \gamma_{1,2} \gamma_{1,1} (\alpha_{\text{crit,max}}(t) - \alpha(t)). \label{eqn:higher-order-alpha-constraint_upepr}
\end{align}
with \(\gamma_{1,1}\) and \(\gamma_{1,2}\) being positive constants. As later used in Eq. \eqref{CBF_function_plant2}, $\psi_{2,2}$ for the HOCBF constraint of the lower state bound on $\alpha(t)$  can be derived very similar to the one of $\psi_{1,2}$. The variables $\dot{\alpha}(t)$ and $\ddot{\alpha}(t)$ are higher order state information of the considered dynamics.  However, since no functional relationship on $\ddot{\alpha}(t)$ is directly available, we propose to estimate the needed higher-order information of the vehicle dynamics based on known dynamics. To do so, we first consider again Eq.~\eqref{eqn:alpha_dot}:
\begin{equation*}
        \dot{\alpha}(t)  = \frac{QS}{mV} \left(C_{Z,\alpha}(M) \alpha(t) +C_{Z,\delta}(M) \delta(t)  \right) + q(t).
\end{equation*}

We differentiate Eq.~\eqref{eqn:alpha_dot} with respect to time. Since \(V\) varies slowly, and with that $Q$ as well, we assume them to be quasi-constant, which yields:
\begin{equation}
    \ddot{\alpha}(t) = \frac{QS}{mV} \left( \frac{\partial C_{z,\alpha}(M)}{\partial t} \alpha(t) + C_{z,\alpha}(M) {\dot{\alpha}}(t) + \frac{\partial C_{z,\delta}(M, \alpha)}{\partial t} \delta + C_{z,\delta}(M) \dot{\delta}(t) \right) + \dot{q}(t).
    \label{eqn:dot_dot_alpha_pre}
\end{equation}
Under consideration of aerodynamic coefficients, which primarily depend on the Mach number, for which we can also assume slowly changing properties relative to the considered attitude dynamics, we assume \cite{Lee2020}:
\begin{equation}
    \frac{\partial C_{z,\alpha}(M)}{\partial t} \approx 0, \quad 
    \frac{\partial C_{z,\delta}(M)}{\partial t} \approx 0.
    \label{eqn:assumtions_aerodynamics}
\end{equation}
With integrating Eq.~\eqref{eqn:assumtions_aerodynamics} back into the Eq.~\eqref{eqn:dot_dot_alpha_pre}, we obtain:
\begin{equation}
    {\ddot{\alpha}}(t) = \frac{QS}{mV} \left( \, C_{z,\alpha}(M) \, {\dot{\alpha}}(t)+ C_{z,\delta}(M)  \dot{\delta}(t)  \, \right) + \dot{q}(t).
    \label{eqn:dot_dot_alpha}
\end{equation}

Following Definitions \ref{higher_order_safety_requirement} and \ref{higher_order_safety_requirement2}, with their necessary inequality conditions, we consider integrating the following angle of attack upper bound inequality constraint in the QP safety filter and aim to satisfy :
\begin{equation}
    \psi_{1,2}(\alpha(t)) = -{\ddot{\alpha}}(t) - (\gamma_{1,1}+  \gamma_{1,2}){\dot{\alpha}}(t) + \gamma_{1,2} \gamma_{1,1} \alpha_{\text{crit,max}}(t) - \gamma_{1,2} \gamma_{1,1}\alpha_r(t) \geq 0, \quad \forall t \geq 0.
    \label{CBF_function_plant}
\end{equation}
Following the same derivation steps as for the upper bound constraint based on \(h_1\), we obtain the following inequality constraint for the lower bound on $\alpha(t)$:
\begin{equation}
{\psi}_{2,2}(\alpha(t)) = {\ddot{\alpha}}(t) + (\gamma_{1,1}+\gamma_{1,2}) {\dot{\alpha}}(t) + \gamma_{2,2} \gamma_{2,1} \alpha_r(t) - \gamma_{2,2} \gamma_{2,1} \alpha_{\text{crit,min}}(t) \geq 0 , \quad \forall t \geq 0.
\label{CBF_function_plant2}
\end{equation}

As discussed in \cite{Xiao_2022hocbf,Agrawal2021}, the construction of valid HOCBFs for systems with relative degree greater than one and limited control authority requires careful tuning of the associated design parameters. The choice of these parameters is essential to ensure that the resulting inequality constraints yield forward-invariant safe sets under actuation constraints. 

{\color{black}
In the current setting, two HOCBF constraints are formulated to enforce upper and lower bounds on the regulated angle-of-attack variable, \(\alpha(t)\). Due to the symmetric nature of the constraints and the similarity of the closed-loop dynamics for both conditions, we adopt a unified tuning configuration such that \(\gamma_{1,1} = \gamma_{2,1} = \gamma_1\) and \(\gamma_{1,2} = \gamma_{2,2} = \gamma_2\).

\begin{figure}[ht]
    \centering
    \includegraphics[width=1\linewidth]{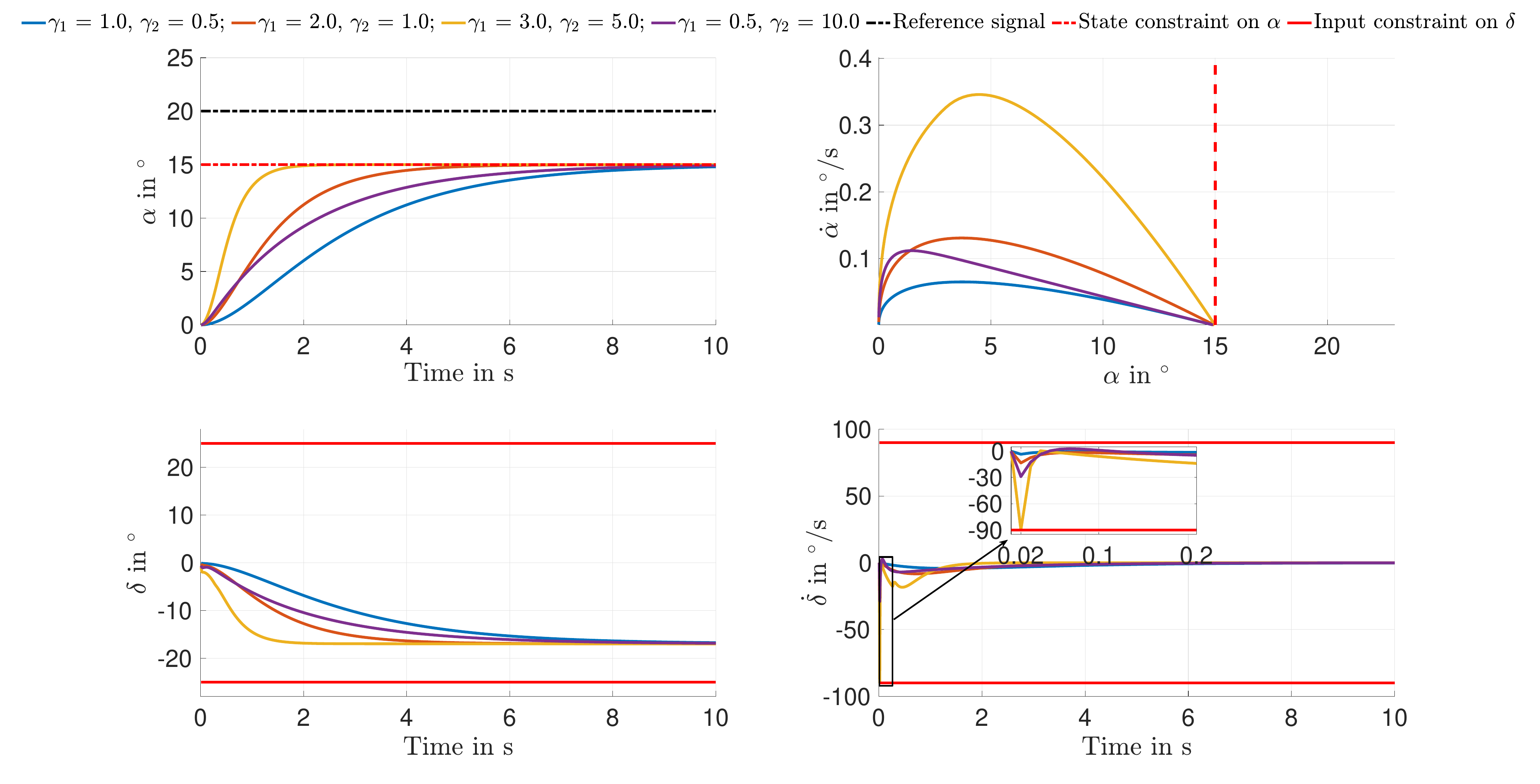}
    \caption{System response for different HOCBF parameter combinations under a reference command of \(\alpha_r = 20^\circ\) and a safety constraint of \(\alpha \leq 15^\circ\).}
    \label{fig:HOCBF_tuning}
\end{figure}

Figure~\ref{fig:HOCBF_tuning} presents simulation results for a scenario where the reference command is an unsafe step input of \(\alpha_r = 20^\circ\), and the safety constraint is given by \(\alpha(t) \leq 15^\circ\). The upper-left subplot shows that all considered HOCBF parameter combinations successfully enforce the state constraint, highlighting the theoretical strength of HOCBFs in guaranteeing safe set invariance, provided that no limiting actuator effects are present.

However, as illustrated in the lower subplots, the different parameter configurations result in different demands on the available actuators. While magnitude constraints are not critical in this scenario, the actuator rate limits influence the feasibility of the safety filter. For instance, the parameter setting \((\gamma_1, \gamma_2) = (3,\ 5)\), shown in yellow, leads to a more aggressive safety enforcement, pushing the control rate \(\dot{\delta}\) at one point to its limit. This behavior is further clarified in the magnified inset (bottom right), showing that the system operates at the edge of feasibility. While such a setting ensures minimal compromise on performance while rendering the system safe, it may not be desirable in practice due to excessive control activity. Therefore, the aim is to identify a setting that balances safety with minimal intervention and reduced actuation effort. In the regarded case, the setting \((\gamma_1, \gamma_2) = (2,\ 1)\) (orange curve) yields a favorable trade-off—preserving constraint satisfaction without saturating the rate limits, suggesting the further use of this parameter setting for the CBF-based safety filter.
}
\section{Numerical Results}
\label{sec:Numerical Results}
A simulation-based analysis was carried out to examine the proposed FEP methodology. The flight dynamics model presented in Sect.~\ref{Sect: Nonlinear_Missile_Model} was implemented in a MATLAB/Simulink environment. 

\subsection{Nominal single operating point FEP performance}
As discussed in Section~\ref{sec:Nonlinear Attitude Controller Design}, classical FEP strategies such as RC or MBCF rely heavily on the underlying controller’s ability to track a reference signal. Hence, the safety guarantees of these approaches are fundamentally dependent on the chosen controller settings. To illustrate this dependency, three control configurations: (i) RC with a conservative controller (blue), (ii) RC with an aggressive controller (green), and (iii) a CBF-based safety filter using the aggressive controller as baseline input (orange) are compared.
\begin{figure}[ht]
    \centering  
    \includegraphics[width=1\linewidth]{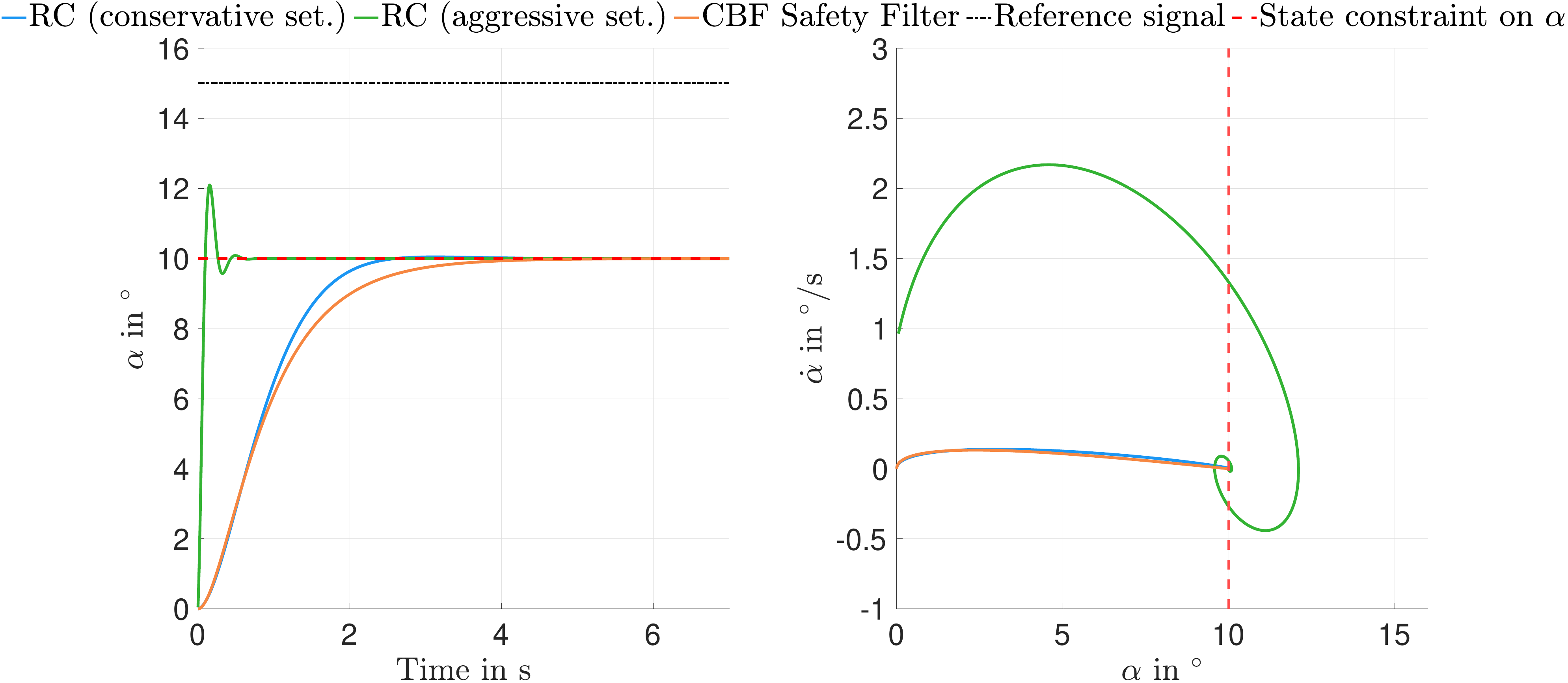}
    \caption{Comparison of three FEP strategies for a commanded angle of attack of \( \alpha_r = 15^\circ \).}
    \label{fig:reference_clipping_comparison}
\end{figure}
Figure~\ref{fig:reference_clipping_comparison} presents simulation results for the three control cases. The left subplot shows the time series of the angle of attack \( \alpha(t) \), where a reference command of \( \alpha_r(t) = 15^\circ \) is applied. For illustration purposes, the upper bound on the angle of attack is artificially constrained to \( \alpha_{\max} = 10^\circ \). The green trajectory in Figure~\ref{fig:reference_clipping_comparison}, representing the aggressive controller, shows significantly faster tracking performance but exhibits overshoot during the transient phase, briefly violating the safety constraint. The blue trajectory representing the closed-loop system with the conservative controller yields a slow reference tracking with no overshoot, suggesting guarantees on FEP when a matching controller setting is found. However, such controller settings will ultimately result in a conservative controller with limited tracking capabilities, also in flight envelope regions where no compromise on control performance is needed for safety reasons. To overcome this issue, a safety filter can be used. The orange trajectory illustrates the outcome of applying a CBF-based safety filter that dynamically adjusts the control input from the aggressive controller, as proposed in Eq.~\eqref{eqn:QP_CBF_FEP}. The integrated safety filter ensures strict constraint satisfaction, independent of the chosen controller, without compromising performance when the state is comfortably within‚ the safe set. The corresponding phase plot on the right of Fig.~\ref{fig:reference_clipping_comparison} further illustrates the described control characteristics of the different controller configurations.

{\color{black}
\subsection{Single Operating Point FEP Performance under Disturbances and Parametric Uncertainties}
While the previous results focused on the nominal performance of FEP strategies, the following part investigates the robustness of CBF-based safety filters in the presence of external disturbances and model uncertainties. It is important to note that although CBFs are a powerful tool that provides formal guarantees on forward invariance, these guarantees fundamentally depend on the accuracy of the underlying system model. In real-world scenarios, however, discrepancies between the model and the actual system behavior, caused by disturbances, degradation, or unmodeled dynamics, can cause the system to violate safety constraints. This limitation is inherent in all model-based techniques. Adaptive CBF approaches, such as those proposed in \cite{Taylor2020-em, Autenrieb2023b}, aim to restore safety guarantees through online adaptation of model parameters. However, the integration of such adaptive schemes is beyond the scope of this study. Instead, during the analysis of the robustness properties, it was observed that standard CBF safety filters are unable to maintain FEP when the system is subject to external disturbances or significant parametric uncertainties. To address this issue, we utilize the inherent connection between CBFs and Control Lyapunov Functions (CLFs), as proposed by \cite{Catellani_2024}. Specifically, in situations where the system leaves the safe set, the barrier function \( h_i(x) \) associated with the respective HOCBF constraint can be reinterpreted as a stabilizing objective by defining an auxiliary CLF:
\begin{equation}
    V_i(x) = 
    \begin{cases}
        0 & \text{if } h_i(x) \geq 0, \\
        -h_i(x) & \text{if } h_i(x) < 0,
    \end{cases}
\end{equation}
where \( i \in \{1,2\} \) refers to the upper and lower bounds on the angle-of-attack \(\alpha\) as defined in Eqs.~\eqref{eqn:h_1} and \eqref{eqn:h_2}. This Lyapunov-like function is integrated into the QP-based filter such that, whenever the system exits the safe set, the controller actively drives the state back toward the safe region via $\dot V_i(x) \leq 0$. The resulting extended CBF safety filter modifies the optimization problem only when \(h_i(x) < 0\), providing a recovery mechanism based on stabilizing feedback. To validate the proposed approach, we first consider a simulation case involving dynamic disturbances. The disturbances are explicitly incorporated into the adjusted nonlinear flight dynamics as additive terms:
\begin{align*}
    \dot{\alpha}_w(t) &= \frac{QS}{mV} \left( C_{Z,\alpha}(M)\alpha(t)+C_{Z,\delta}(M)\delta (t)\right) + q(t) + w_{\dot \alpha},\\
    \dot{q}_w(t) &= \frac{QSd}{I_{yy}} \left(C_{M,\alpha}(M)\alpha(t)+C_{M,q}(M)\frac{d}{2V}q(t)+C_{M,\delta}(M)\delta(t) \right)+ w_{\dot q}, \\
    n_{z,w}(t) &= \frac{QS}{mg} \left(C_{Z,\alpha}(M)\alpha(t)+C_{Z,\delta}(M)\delta(t)\right) + w_{n,z},
\end{align*}
where the disturbances are chosen as \( w_{n_z} = 0.3 \), \( w_{\dot q} = 5.7296^\circ/s^2 \), and \( w_{\dot \alpha} = 17.1887^\circ/s \).

\begin{figure}[ht]
    \centering
    \includegraphics[width=1\linewidth]{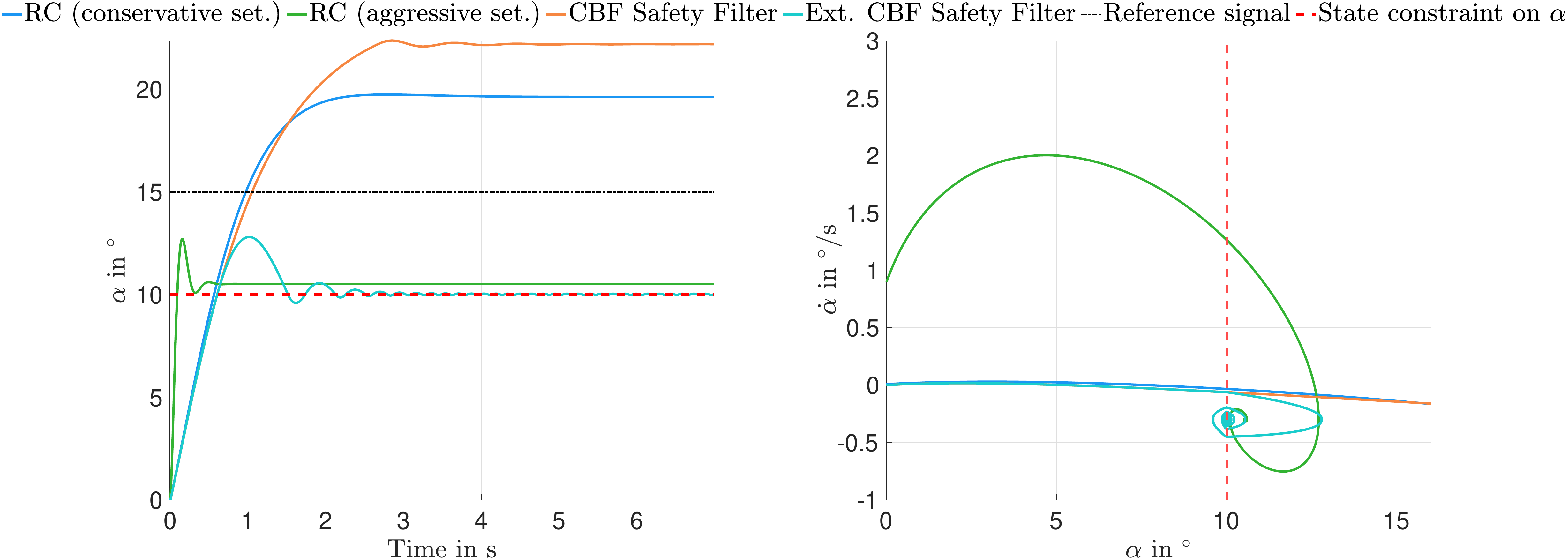}
    \caption{Comparison of FEP strategies under a reference command of \( \alpha_r = 15^\circ \) in the presence of dynamic disturbances.}
    \label{fig:fep_disturbance}
\end{figure}

Figure~\ref{fig:fep_disturbance} presents the resulting system behavior. As shown in the left subplot, the standard CBF filter (orange) and the baseline reference clipping controllers (green and blue) fail to enforce the state constraint in the presence of disturbances. In contrast, the extended CBF filter with CLF backup (cyan) momentarily leaves the safe set but actively drives the system back towards the constraint boundary, restoring safety. This behavior empirically confirms the stabilizing effect of the CLF augmentation and validates the theoretical connection between CBFs and CLFs for safety recovery.

While disturbances primarily account for real-time deviations from the nominal trajectory, parametric uncertainties represent structural mismatches between the modeled and actual system behavior. These may arise from outdated aerodynamic data, actuator modeling errors, or changes in environmental conditions. To model parametric uncertainties, static deviations in key aerodynamic coefficients are considered as: $\Delta C_{Z,\alpha}=-40\%$,$\Delta C_{Z,\delta}=-30\%$, and $\Delta C_{m,q}=-10\%$.

\begin{figure}[ht]
    \centering
    \includegraphics[width=1\linewidth]{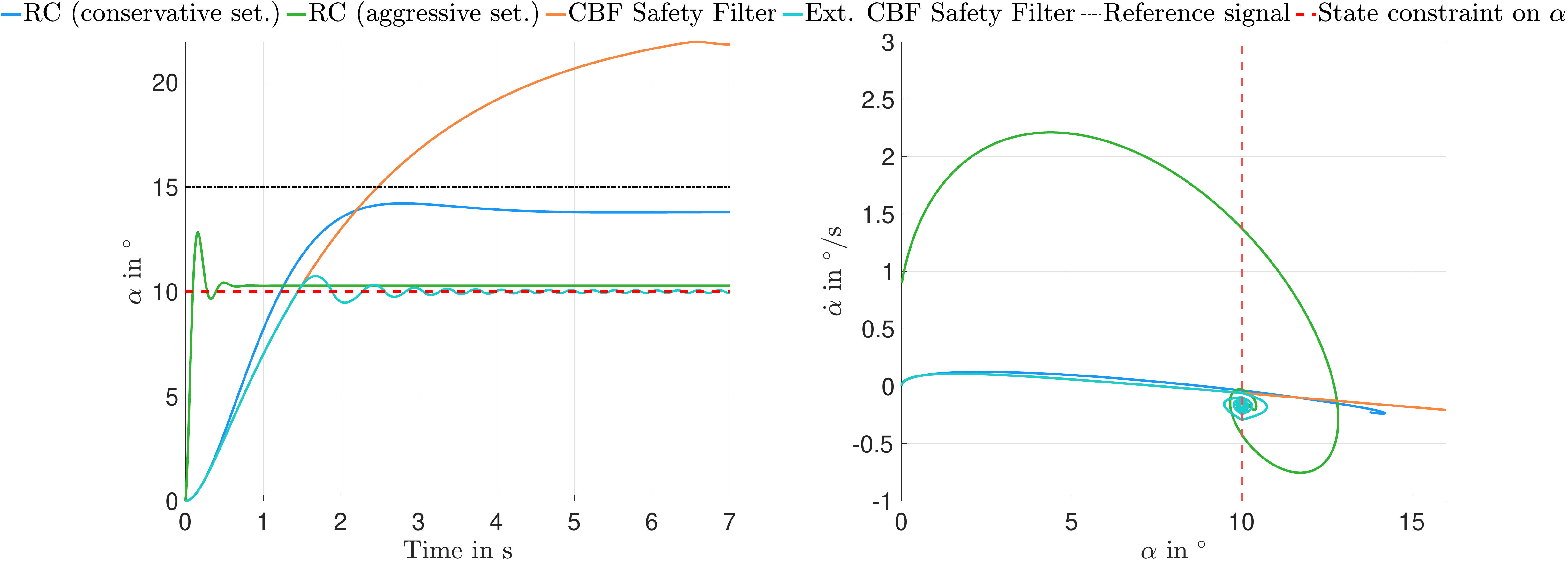}
    \caption{Comparison of FEP strategies for a commanded angle of attack of \( \alpha_r = 15^\circ \) in the presence of parametric uncertainties.}
    \label{fig:fep_uncertainty}
\end{figure}

Figure~\ref{fig:fep_uncertainty} illustrates the resulting system trajectories under parametric uncertainty. As in the disturbance case, all baseline controllers and the standard CBF filter fail to maintain the state within the safety bounds. However, the extended CBF filter with CLF-based recovery (cyan) again enables the system to re-enter the safe set after an initial violation. This demonstrates that, even under significant model mismatches, the proposed extension enhances the robustness of CBF-based FEP strategies by providing graceful safety recovery as long as sufficient control authority is available.
}

\subsection{Extended FEP performance analysis}
To further validate the proposed approach across a wide range of operating conditions, an extended simulation is conducted over a 150-second trajectory covering a flight velocity range from \( V = 100\, \text{m/s} \) to \( V = 1500\, \text{m/s} \). In this scenario, the complete flight envelope is explored while both stall and load factor constraints on the angle of attack are considered, as discussed in Section~\ref{sec:he Flight Envelope Protection Problem}. 
\begin{figure}[ht]
    \centering
    \includegraphics[width=1\linewidth]{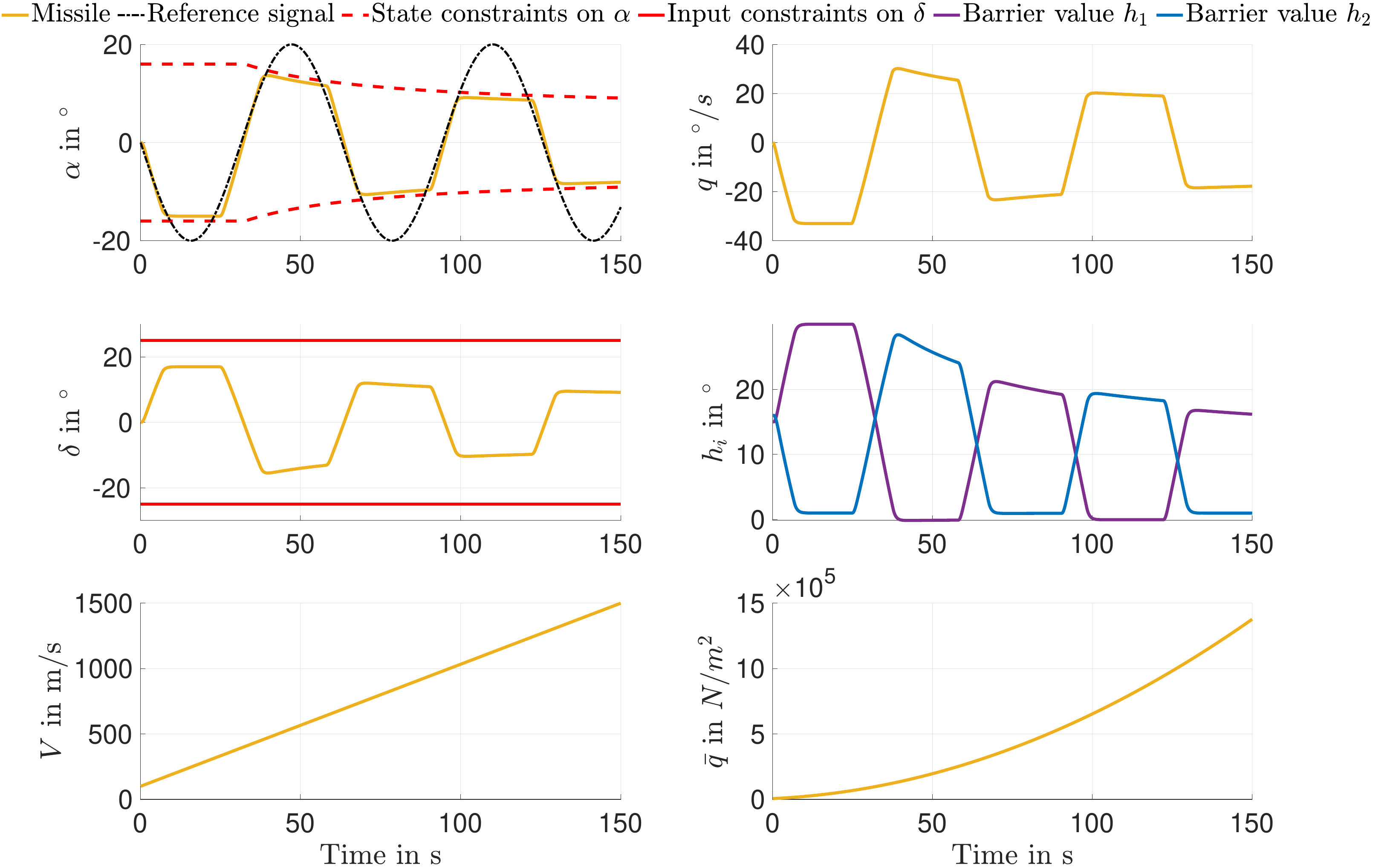}
    \caption{Closed-loop response using a CBF-based safety filter over an extended maneuver covering the entire flight envelope.}
    \label{fig:envelope_case_extended}
\end{figure}
Figure~\ref{fig:envelope_case_extended} presents the results of this case. As illustrated in the upper left plot of Figure~\ref{fig:envelope_case_extended}, the applied reference command \( \alpha_r(t) \) (black dash-dotted line) is sinusoidal and intentionally violates the imposed safety constraints, including the stall limit of \( \pm 15^\circ \) and load factor constraints of \( \pm 10 \). Initially, the stall constraints are most critical due to low dynamic pressure, while load factor limitations dominate later phases as the vehicle accelerates. Despite these challenges, the proposed CBF-based safety filter successfully ensures strict satisfaction of the considered state constraints throughout the entire flight envelope. The lower subplots of Figure~\ref{fig:envelope_case_extended} show the time evolution of additional flight variables. Notably, both barrier values, $h_1$ and $h_2$, remain strictly positive, confirming the forward invariance of the safe set. The actuator input \( \delta(t) \) remains within admissible bounds, illustrating that the safety filter accounts for input constraints while maintaining closed-loop performance. 


\section{Conclusion}
\label{sec:Conclusion}
This paper introduced a new perspective on addressing the FEP problem using an optimization-based CBF approach. Conventional FEP methods were shown to be insufficient in guaranteeing constraint satisfaction due to their reliance on tracking performance rather than explicit constraint enforcement. To address this, the FEP problem was formulated as a state-constrained control problem, and a CBF-based safety filter was designed to ensure strict adherence to safety constraints while maintaining control performance. The proposed framework was integrated into a nonlinear flight control system, leveraging HOCBFs to handle higher relative degree constraints. This allowed for the enforcement of critical flight envelope limits, including load factor and angle-of-attack constraints, while minimizing conservatism. Numerical simulations demonstrated the effectiveness of the approach in dynamically enforcing safety constraints, outperforming traditional RC methods, particularly in transient phases where constraint violations are typically observed. To further enhance robustness against disturbances and parametric uncertainties, the proposed framework was extended with a CLF-based recovery mechanism that actively drives the system back into the safe set whenever temporary constraint violations occur. The results highlight the potential of the CBF-based FEP systems to enhance safety while allowing aerospace systems to operate closer to their performance limits. Future work will focus on extending the approach to more complex flight scenarios, including actuator dynamics, and on real-time implementation for experimental validation.

\section*{Acknowledgments}
The author would like to thank Dr.-Ing. Mark Spiller from the DLR Institute of Flight Systems, Department of Unmanned Aircraft, for fruitful conversations and valuable suggestions during the early stages of this manuscript. The author also thanks Dr. Anuradha Annaswamy and Peter A. Fisher from the Active Adaptive Control Laboratory at MIT for stimulating discussions on control barrier functions, which provided valuable perspectives during the development of this work.
\bibliography{sample}

\end{document}